\documentclass[12pt]{article}

\usepackage{amsmath,amssymb,tensor, footnote,caption,cite}

\usepackage[unicode,linktocpage=true,plainpages=false,pdfpagelabels=false]{hyperref}

\newcommand{\pp}{\varphi}
\newcommand{\de}{\delta}
\newcommand{\dd}{\partial}
\newcommand{\te}{\theta}
\newcommand{\al}{\alpha}
\newcommand{\ga}{\gamma}
\newcommand{\m}{\mu}
\newcommand{\n}{\nu}
\newcommand{\ta}{\tau}
\newcommand{\be}{\beta}
\newcommand{\can}{\mathcal{T}}
\begin{document}

\title{Noether and Belinfante stress-energy tensors for theories with arbitrary Lagrangians of tensor fields}
\author{
R.~V.~Ilin\thanks{E-mail: st030779@student.spbu.ru},
S.~A.~Paston\thanks{E-mail: pastonsergey@gmail.com}\\
{\it Saint Petersburg State University, Saint Petersburg, Russia}
}
\date{\vskip 15mm}
\maketitle

\begin{abstract}
We investigate the connection between stress-energy tensor (SET) arising from Noe\-ther's theorem and Belinfante SET which can be obtained as a right-hand side of the Einstein's equation in the flat metric limit. This question is studied in the wide class of Poincar\`e-invariant field theories with actions which depend on the tensor fields of arbitrary rank and their derivatives of arbitrary order. For this class we derive the relation between these SET and present the exact expression for the difference between them. We also show that the difference between corresponding integrals of motion can be expressed as a surface integral over 2-dimensinal infinitely remote surface.
\end{abstract}

\newpage

\section{Introduction}
It is well known that a wide class of field theories satisfies Noether's theorem which states that there is a correspondence between global symmetries of the action and conserving integrals of motion \cite{Noether}. For example, translation invariance of the theory corresponds to the energy-momentum vector. In the simplest case of scalar field $\pp$ with action depending on the fields and their first derivatives the density of this vector have the following form:
\begin{equation}\label{q1}
\can^\mu{}_\nu=\frac{\dd L}{\dd\dd_\mu \pp}\dd_{\nu}\pp-\de^{\mu}_\nu L;\qquad \dd_{\mu}\can^\mu{}_{\nu}=0,
\end{equation}
where $L$ is the Lagrangian density. This quantity is called Noether or "canonical" SET.\par

The integrals of motion arising from Noether's theorem are usually observable and have clear physical meaning. However, in general their densities do not adopt these properties. The good example of this issue is an ambiguity in the definition of gravitational energy density, the well known problem of non-localizability of energy in gravity \cite{faddeev-UFN1982}. Another possible obstacle connected with Noether SET is that, generally speaking, it is not symmetric. This fact leads to a lack of the simple relation between it and angular momentum tensor{} (see \cite{Weinberg}, \S7.4). Due to all these problems, one also use another SET defined as the following:
\begin{equation}
T_{\mu\nu}=-\frac{2}{\sqrt{-g}}\frac{\delta S}{\delta g^{\mu\nu}}\Bigg|_{g_{\mu\nu}=\eta_{\mu\nu}};\qquad \dd_\mu T^{\mu\nu}=0,
\label{metric}
\end{equation}
where $S$ -- action of matter coupled with gravity, $g_{\mu\nu}$ -- spacetime metric, $\eta_{\mu\nu}$ -- Minkowski metric. This tensor is called Belinfante SET (Hilbert SET or metric SET). It appears to be symmetric by definition and observable because it is exactly the right-hand side of Einstein's equation. It is worth noting that the correspondent conservation law follows from the satisfaction of the matter equations of motion and from the prescription of general covariance of the action of matter coupled with gravity.
\par

In the papers \cite{belifante,Rosenfeld} it was noted for the first time that Belifante and canonical SET are connected by the following relation:
\begin{equation}
\mathcal{T}^{\mu\nu}+\dd_{\alpha}B^{\alpha\mu\nu}=T^{\mu\nu},\qquad B^{\alpha\mu\nu}=-B^{\mu\alpha\nu}.
\label{eqviw}
\end{equation}
It should be noted that the structure of their difference guarantees the fact that the correspondent integrals of motion differ from each other by the surface integral over an infinitely remote 2D spacelike surface.
In the papers mentioned above the authors proved this relation for theories with first derivative Lagrangians. In the following years the relation between SET have been generalized for second derivative Lagrangians (for example \cite{Ohanian}; see also \cite{Mickevich1969}) and also for  first derivative Lagrangians with arbitrary transformation properties of the fields \cite{GoMa}.\par
The present paper investigate the relation between Noether SET and Belinfante SET for Poincare-invariant theories with actions depending on tensor fields and their derivatives up to an arbitrary order $n$. We will show that in this case the relation \eqref{eqviw} still holds true but the additional term $B^{\alpha\mu\nu}$ is no longer antisymmetric over first two indices. Then we will show that despite this fact, the correspondent integrals of motion turn out to differ from each other by a surface term.

\section{The proof of the relation between Noether and Belifante SET}
Consider the Poincar\`e-invariant theory of the fields $\pp_A$ (for shortness we will use multi-index $A\equiv\mu_1...\mu_n$) with Lagrangian density depending on fields and its derivatives up to the $n$th order:
\begin{equation}
L(\pp_A,\dd_{\alpha_1}\pp_A,...,\dd_{\al_1}...\dd_{\al_n}\pp_A).
\label{lagr}
\end{equation}
Since the Lagrangian density $L$ has no explicit dependence on coordinates, this theory is invariant under spacetime translations. Hence, as it was mentioned in the Introduction, there are 4 integrals of motion which are defined by $0\mu$-components of Noether SET $\can^\tau{}_\theta$. It's easy to show that for theory with Lagrangian \eqref{lagr} the expression for Noether SET can be written as follows:
\begin{equation}
\mathcal{T}^\tau{}_\te=\sum_{j=1}^{n}\sum_{i=1}^{j}(-1)^{i+j}
\left(\partial_{\tau_{i+1}}...\partial_{\tau_{j}}\frac{\partial{L}}{\partial(\partial_{\tau}\partial_{\tau_2}...\partial_{\tau_j}\varphi_A)}\right)
\partial_{\tau_2}...\partial_{\tau_i}\partial_{\theta}\varphi_A-{L}\delta^{\tau}_{\theta}.
\label{NoetherT}
\end{equation}
Hereafter we assume that $\dd_{\alpha_k}...\dd_{\alpha_m}$ equals to $\dd_{\al_k}$ if $m=k$, equals to $1$ if $m=k-1$  and equals to $0$ if $m<k-1$. \par

Noether's theorem proof (i.e. proof of the fact that \eqref{NoetherT} satisfies differential conservation law \eqref{q1}) is based on the consideration of  Lagrangian variation with respect to global infinitesimal translation. Application of this method in case of curved spacetime, where translational symmetry has become local, can lead to some new useful equations. In particular, one of these equations is somewhat analogous to formula \eqref{eqviw} in the limit $g_{\mu\nu}\to\eta_{\mu\nu}$. So, first of all, one needs to define the so called form variation with respect to infinitesimal local coordinate transformation $x^{\mu}\rightarrow x'^{\mu}=x^{\mu}+\xi^{\mu}(x)$:
\begin{equation}
\bar{\de}q(x)=q'(x)-q(x).
\end{equation}
Its easy to show that form variation of tensor field $\pp_A$ is equal to
\begin{equation}
\bar{\delta}\varphi_A=-\xi^{\alpha}\partial_{\alpha}\varphi_A-(\partial_{\mu_1}\xi^{\alpha})\varphi_{\alpha\mu_2...\mu_k}-
(\partial_{\mu_2}\xi^{\alpha})\varphi_{\mu_1\alpha\mu_3...\mu_k}-\ldots.
\label{var}
\end{equation}
Note that this equation is not valid for spinor fields and this case requires specific approach.
For further consideration it is also useful to calculate the form variation of metric $g_{\mu\nu}$:
\begin{equation}
\bar{\delta}g_{\m\n}=-D_{\m}\xi_{\n}-D_{\n}\xi_{\m}\label{Variation gmu}
\end{equation}
and form variation of a scalar density $\hat L$ which can be written as $\sqrt{-g}L$ {} (here $g$ is a determinant of $g_{\mu\nu}$ and $L$ is a scalar):
\begin{equation}
\bar{\delta}\hat L=-\partial_{\alpha}(\xi^{\alpha}\hat L).
\label{ScalarDensity}
\end{equation}\par
Now we come directly to the proof of the formula analogous to \eqref{eqviw}.
As it was mentioned above, one needs to include certain minimal coupling with gravity in theory \eqref{lagr}. As a result, new Lagrangian density depends on the following arguments:
\begin{equation}
\hat L(\pp_A, \dd_{\tau_1}\pp_A,\ldots,\dd_{\tau_1}...\dd_{\tau_n}\pp_A,g_{\mu\nu}, \dd_{\tau_1}g_{\mu\nu},\ldots,\dd_{\tau_1}...\dd_{\tau_n}g_{\mu\nu}).
\label{Lnew}
\end{equation}
Since it is a scalar density, its form variation is given by \eqref{ScalarDensity}. On the other hand, the form variation $\bar{\delta}$ has the properties of differentiation and hence $\bar{\delta}\hat L$ can be expressed as a linear combination of  \eqref{var},\eqref{Variation gmu} and their derivatives. Coefficients of this combination are the derivatives of $\hat L$ with respect to $\varphi_A$, $g_{\m\n}$ and their derivatives. Using this fact together with equations of motion and the definition of Belinfante SET\eqref{metric} and taking the limit $g_{\mu\nu}\to\eta_{\mu\nu}$ in the appearing equation one can obtain the following:
\begin{equation}
\begin{split}
\partial_{\tau}\sum_{j=1}^{n}\sum_{i=1}^{j}(-1)^{i+j}\Biggl[
\left(\partial_{\tau_{i+1}}...\partial_{\tau_{j}}\frac{\partial L}{\partial(\partial_{\tau}\partial_{\tau_2}...\partial_{\tau_j}\varphi_{A})}\right)
\partial_{\tau_2}...\partial_{\tau_i}\Bigl(\xi^{\alpha}\partial_{\alpha}\varphi_A+(\partial_{\mu_1}\xi^{\alpha})\varphi_{\alpha\mu_2...\mu_k}+\ldots\Bigr)+\\
+2\left(\partial_{\tau_{i+1}}...\partial_{\tau_{j}}\frac{\partial \hat L}{\partial(\partial_{\tau}\partial_{\tau_2}...\partial_{\tau_j}g_{\mu\nu})}\right)
\Bigg|_{g_{\ga\de}=\eta_{\ga\de}}
\partial_{\tau_2}...\partial_{\tau_i}\dd_{\mu}\xi_{\nu}\Biggr]-\partial_{\alpha}(\xi^{\alpha} L)-T^{\mu\nu}\dd_{\mu}\xi_{\nu}=0.
\label{Riemanneq}
\end{split}
\end{equation}
Since $\xi^{\mu}$ and its derivatives are linearly independent at the every point, the Equation \eqref{Riemanneq} is equivalent to a system of $n+2$ equations which can be found by setting to zero the coefficients of $\xi^{\mu}$ and its derivatives.\par
It's easy to show that first equation in this system is exactly the differential conservation law \eqref{q1} for Noether SET \eqref{NoetherT}. Its appearance in system arising from \eqref{Riemanneq} is not unexpectable because the application of Noether's theorem  leads to the particular variant of \eqref{Riemanneq} for $\xi^{\mu}=const$. The second equation in this system leads to the abovementioned relation between Noether's and metric SET:
\begin{equation}\label{sp11}
T^{\be\alpha}-\mathcal{T}^{\be\alpha}=\dd_{\ta}B^{\ta\be\al},
\end{equation}
where
\begin{equation}
\begin{split}
B^{\ta\be\al}=\sum_{j=1}^{n}\Biggl[\sum_{i=1}^{j}(-1)^{i+j}
\left(\partial_{\tau_{i+1}}...\partial_{\tau_{j}}\frac{\partial L}{\partial(\partial_{\tau}\partial_{\tau_2}...\partial_{\tau_j}\varphi_A)}\right)
\Bigl((i-1)\delta^{\be}_{\tau_2}\partial_{\tau_3}...\partial_{\tau_i}\partial^{\alpha}\varphi_A+\\
+\delta_{\mu_1}^{\be}\partial_{\tau_2}...\partial_{\tau_i}\varphi_{\alpha\mu_2...\mu_k}+\ldots\Bigr)+
2(-1)^{j+1}\left(\partial_{\tau_{2}}...\partial_{\tau_{j}}
\frac{\partial \hat L}{\partial(\partial_{\tau}\partial_{\tau_2}...\partial_{\tau_j}g_{\be\al})}\right)\Bigg|_{g_{\ga\de}=\eta_{\ga\de}}
\Biggr].
\label{2nd}
\end{split}
\end{equation}
This quantity is similar to one in \eqref{eqviw} with only difference that it is not necessarily anti-symmetric over first two indices. Hence one can not guarantee that difference between the   energy-momentum vectors corresponding to $T^{\mu\nu}$ and $\mathcal{T}^{\mu\nu}$ equals to the surface term except the case with $n=1$. Nevertheless, one can use the rest of the equations in system \eqref{Riemanneq} in order to show that $B^{00\al}$ admits the following representation:
\begin{equation}\label{sp19}
B^{00\al}=-\dd_k \sum_{l=2}^n (-\dd_0)^{l-2}N_{(l)}^{k\al},
\end{equation}
where an explicit (though quite cumbersome) expression for $N_{(l)}^{k\al}$ can be written using the left-hand side of $l$-th equation in the system that was mentioned after \eqref{Riemanneq}. Therefore the difference between the energy-momentum tensors which can be calculated from $\mathcal{T}^{\mu\al}$ and $T^{\mu\al}$ respectively by integration of its $0\al$-component over $3$-dimensional surface $t=const$ is equal to the surface integral over 2-dimensional infinitely remote surface:
\begin{equation}
P^{\al}-\mathcal{P}^{\al}=\int d^3x\,  \dd_{\ta}B^{\ta0\al}=\int\limits_M dS_k\left[ B^{k0\alpha}+ \sum_{l=2}^n(-\dd_0)^{l-1}N_{(l)}^{k\al}\right],
\label{int}
\end{equation}
where we used \eqref{sp19} and Gauss's theorem.

\section{Conclusion}
We have shown that in Poincar\`e-invariant field theory  with $n$th derivative Lagrangian the Belinfante SET $T^{\mu\nu}$ and the Noether SET $\mathcal{T}^{\mu\nu}$ are connected with each other by the formulae \eqref{sp11}, \eqref{2nd}. While $B^{\al\m\n}$ is anti-symmetric in the case of $n=1$ which is disscused in details in the original works \cite{belifante,Rosenfeld}, in the most general case of arbitrary $n$ the quantity $B^{\al\m\n}$ doesn't have this property. Instead of it, a weaker condition \eqref{sp19} is satisfied which nevertheless leads to the similar result: the integrals of motion are related to each other through a surface term, see ~\eqref{int}.\par
The relation obtained can be used for proceeding from the definition of energy density in the canonical formalism, which is closely connected with quantum theory, to its definition through Belinfante SET, which can be observed by its coupling with gravity. This can be helpful in a number of situations where application of the definition of energy density is needed.\par
The result can be also useful in the searching of correct definition of gravitational energy in the modified theories of gravity. This question was studied in \cite{statja46} for the so-called embedding theory \cite{regge,deser,statja18,statja24}. It was shown that for a variant of embedding theory as a field theory in flat space with high order derivatives \cite{statja25} the definition of gravitational energy as a Belinfante SET gives non-trivial result even for solutions of Einstein's equations. At the same time, Noether's SET appears to be zero for these physically interesting solutions.

{\bf Acknowledgements.}
The work of one of the authors (R.~V.~Ilin) was supported by RFBR grant N~18-31-00169.

%\providecommand{\eprint}[1]{\href{http://arxiv.org/abs/#1}{\texttt{#1}}}
%\bibliographystyle{../../my3beznazv}
%\bibliography{../../paston-grav-e}

\begin{thebibliography}{10}
\newcommand{\enquote}[1]{``#1''}
\providecommand{\url}[1]{\texttt{#1}}
\providecommand{\urlprefix}{URL }
\expandafter\ifx\csname urlstyle\endcsname\relax
  \providecommand{\doi}[1]{doi:\discretionary{}{}{}#1}\else
  \providecommand{\doi}{doi:\discretionary{}{}{}\begingroup
  \urlstyle{rm}\Url}\fi
\providecommand{\eprint}[1]{\href{http://arxiv.org/abs/#1}{\texttt{#1}}}

\bibitem{Noether}
E.~Noether, \emph{Gott. Nachr.} (1918), 235--257, \eprint{arXiv:physics/0503066}.

\bibitem{faddeev-UFN1982}
L.~D. Faddeev,
  \href{http://dx.doi.org/10.1070/PU1982v025n03ABEH004517}{\emph{Soviet Physics
  Uspekhi}}, \textbf{25}: 3 (1982), 130.

\bibitem{Weinberg}
S.~Weinberg, \enquote{The Quantum Theory of Fields}, 1, Cambridge University
  Press, 1995.

\bibitem{belifante}
F.J. Belinfante, \emph{Physica}, \textbf{6} (1939), 887.

\bibitem{Rosenfeld}
L.~Rosenfeld, \emph{Mem. Acad. Roy. Belg. Sci}, \textbf{XVIII}: 6 (1940), 1536.

\bibitem{Ohanian}
Hans~C. Ohanian, \enquote{The Energy-Momentum Tensor in General Relativity and
  in Alternative Theories of Gravitation, and the Gravitational vs. Inertial
  Mass}, 2010, \eprint{arXiv:1010.5557}.

\bibitem{Mickevich1969}
N.~V. Mitskevich, \enquote{Physical fields in General Relativity [Fizicheskie
  polia v obshchei teorii otnositel'nosti]}, Nauka, 1969, in Russian.

\bibitem{GoMa}
M.J. Gotay, J.E. Marsden, \emph{Contemporary Mathematics}, \textbf{132} (1992),
  367--392.

\bibitem{statja46}
D.~A. Grad, R.~V. Ilin, S.~A. Paston, A.A. Sheykin,
  \href{http://dx.doi.org/10.1142/S0218271817501887}{\emph{Int. J. Mod. Phys.
  D}}, \textbf{27} (2018), 1750188, \eprint{arXiv:1707.01074}.

\bibitem{regge}
T.~Regge, C.~Teitelboim, \enquote{General relativity \`a la string: a progress
  report}, in \emph{Proceedings of the First Marcel Grossmann Meeting, Trieste,
  Italy, 1975}, edited by R.~Ruffini, 77--88, North Holland, Amsterdam, 1977,
  \eprint{arXiv:1612.05256}.

\bibitem{deser}
S.~Deser, F.~A.~E. Pirani, D.~C. Robinson,
  \href{http://dx.doi.org/10.1103/PhysRevD.14.3301}{\emph{Phys. Rev. D}},
  \textbf{14}: 12 (1976), 3301--3303.

\bibitem{statja18}
S.~A. Paston, V.~A. Franke,
  \href{http://dx.doi.org/10.1007/s11232-007-0134-9}{\emph{Theor. Math.
  Phys.}}, \textbf{153}: 2 (2007), 1582--1596, \eprint{arXiv:0711.0576}.

\bibitem{statja24}
S.~A. Paston, A.~N. Semenova,
  \href{http://dx.doi.org/10.1007/s10773-010-0456-5}{\emph{Int. J. Theor.
  Phys.}}, \textbf{49}: 11 (2010), 2648--2658, \eprint{arXiv:1003.0172}.

\bibitem{statja25}
S.~A. Paston, \href{http://dx.doi.org/10.1007/s11232-011-0138-3}{\emph{Theor.
  Math. Phys.}}, \textbf{169}: 2 (2011), 1611--1619, \eprint{arXiv:1111.1104}.

\end{thebibliography}

\end{document}